# Single-photon emitters based on NIR colour centres in diamond coupled with solid immersion lenses

D. Gatto Monticone[1,2,3], P. Traina[4], J. Forneris[1,2,3], M. Levi[1], E. Moreva[4], E. Enrico[4], A. Battiato[1,2,3], F. Picollo[2,1,3], G. Brida[4], I. P. Degiovanni[4], P. Olivero[1,2,3]*, M. Genovese[4], G. Amato[4], L. Boarino[4]

[1] *Physics Department, University of Torino, Italy*

[2] *Istituto Nazionale di Fisica Nucleare (INFN), sez. di Torino, Italy*

[3] *Consorzio Nazionale Interuniversitario per le Scienze fisiche della Materia (CNISM), Italy*

[4] *Istituto Nazionale di Ricerca Metrologica (INRiM), Italy*

\* corresponding author:   e-mail: paolo.olivero@unito.it

          ph: +39 011 670 7366

          fax: +39 011 670 7020

**Abstract**

Single-photon sources represent a key enabling technology in quantum optics, and single colour centres in diamond are a promising platform to serve this purpose, due to their high quantum efficiency and photostability at room temperature. The widely studied nitrogen-vacancy centres are characterized by several limitations, thus other defects have recently been considered, with a specific focus of centres emitting in the Near Infra-Red. In the present work, we report on the coupling of native near-infrared-emitting centres in high-quality single-crystal diamond with Solid Immersion Lens structures fabricated by Focused Ion Beam lithography. The reported improvements in terms of light collection efficiency make the proposed system an ideal platform for the development of single-photon emitters with appealing photophysical and spectral properties.

**1. Introduction**

Single-photon sources represent a key enabling technology in quantum optics and quantum technologies [1-7]. Diamond is a promising platform for the development of single-photon emitters, due to a remarkable variety of luminescent defects [8, 9] with appealing emission properties combined with the broad spectral transparency of the hosting crystal structure. The negatively charged nitrogen-vacancy ($NV^-$) complex had a prominent role in the development of the first diamond-based single photon emitters [10-12] due to its abundance, quantum efficiency and well understood electronic transition structure [13]. In the last decade, quantum cryptography schemes [14-17] as well as more fundamental measurements of quantum complementarity and entanglement [18-21] were demonstrated with $NV^-$ emitters. Recently, new defect types were investigated with the purpose of overcoming some limitations of the $NV^-$ centre, namely its strong phonon coupling, relatively long lifetime and charge-state blinking. In particular, recent works reported on centres emitting in the Near-Infra-Red (NIR), which would offer a higher degree of compatibility with the optimal spectral efficiency of Si-based photodetectors. To this purpose, different NIR-emitting centres were explored with regards to their fabrication strategies, physical properties and functional performances [22-25] and their suitability for the implementation of quantum communication schemes was assessed [26, 27]. In particular, the role of high-temperature (T > 1000 °C) annealing processes on the activation of NIR-emitting centres in pristine high-quality single-crystal diamonds was recently investigated, both in terms of formation yield and spectral emission properties [28, 29].

In the above-mentioned works, the NIR-emitting centres display appealing photo-physical properties such as narrow (i.e. few nanometers) zero-phonon-line (ZPL) emissions in the $\lambda = 740$–$780$ nm spectral range, ~ns radiative lifetimes, full linear polarization and high emission rates (i.e. $\sim 5 \times 10^5$ counts $s^{-1}$ in saturation conditions, about 1 order of magnitude higher than in $NV^-$ centres). In particular, the emission rate is a crucial functional parameter in application where high single-photon fluxes are needed, such as quantum cryptography.

While NIR colour centres can be produced by suitably activated impurities in the diamond lattice, an unambiguous attribution of their structure would pave the way to two main fabrication strategies, i.e. i) the incorporation of given impurities from either the plasma or the substrate during chemical vapour deposition (CVD), or ii) the direct implantation of the same impurities into bulk crystals [22]. The latter approach offers the advantage of allowing a higher degree of control and spatial resolution in the formation of the centres, although centres created by ion implantation are generally characterized by significantly lower emission rates with respect to their as-grown counterparts, which are typically formed in isolated micro/nano-crystals.

Moreover, colour centres located in the sample bulk, although being insensitive to surface defects, suffer from an intrinsic limitation in their light collection efficiency, since the high value of the refractive index in diamond causes a great portion of emitted light to be internally reflected by the diamond/air interface. The semi-aperture $\theta$ of the light cone that can be transmitted from diamond to air (or into a different medium, such as in oil-immersion lenses) is, in paraxial approximation, equal to:

$$\theta = \sin^{-1}\left(\frac{n_0}{n_d}\right) \quad (1)$$

where $n_0$ is the refractive index of air (1) or oil (~1.52) and $n_d$ is the refractive index of diamond (~2.41). For a diamond/air interface $\theta = 25°$ and for a diamond/oil interface $\theta = 39°$, significantly different from the typical acceptance of a standard objective (i.e. ~64°).

The solid angle $\Omega$ of a cone with a semi-aperture $\theta$ is given by:

$$\Omega = 2\pi(1-\cos\theta) \quad (2)$$

Therefore, in the case of an isotropic emitter in the bulk of the sample, a standard air-objective only collects ~16% of the light that would be collected by a hypothetical corresponding emitter free-standing in air. Similarly, an oil immersion lens only collects ~39% of the light emitted by a corresponding emitter in oil.

An effective solution to the collection efficiency issue is represented by the direct fabrication of Solid Immersion Lenses (SILs) into the diamond crystal. SILs are widely known in integrated optics, as their operating principle is straightforward: the SIL is a portion of a sphere whose geometrical centre is positioned in correspondence of the emitter location. As shown in Fig. 1, in this way a significant portion of the light rays coming towards the surface encounters the diamond/air interface in normal direction, filling more effectively the acceptance cone of the objective. SILs were fabricated in diamond by means of Focused Ion Beam (FIB) lithography, and significant improvements in light collection efficiency from $NV^-$ centres were reported [30-34].

In the present work we report on the direct FIB fabrication of a SIL in single-crystal diamond registered to a NIR-emitting single colour centre, demonstrating a significant increase in light collection efficiency from these appealing single-photon emitters.

## 2. Optical characterisation setup

A $3\times3\times0.3$ mm$^3$ single-crystal CVD diamond sample produced by Element Six was employed. The sample is classified as "optical grade" type IIa, its nominal substitutional nitrogen and boron concentrations being <1 ppm and <0.05 ppm, respectively. The crystal orientation is <100> and it is optically polished (nominal surface mean roughness $R_a$ < 30 nm) on both of its two larger faces.

The as-grown sample was processed with a 1-hour high-temperature (T = 1450 °C) annealing in vacuum to induce the formation/activation of luminescent centres with good emission properties with an average surface density of ~1-10 centres per $50\times50$ µm$^2$ and a uniform distribution within the range of depths probed by the confocal system (~50 µm) [29]. After annealing, the sample was cleaned by oxidation, firstly with a 30-min-long 600 °C annealing in air and subsequently by exposure to an oxygen plasma (microwave power: 30 W, $O_2$ flow: 30 sccm, pressure in chamber: 0.25 mbar, duration: 30 min), with the scope of removing partially graphitized regions from the sample surface.

Optical characterization was performed with a single-photon-sensitive confocal photoluminescence (PL) microscopy system, as schematically shown in Fig. 2. The NIR centres were excited using a continuous laser light at $\lambda$ = 690 nm emitted by a solid-state laser source.

The excitation beam was coupled into a single-mode optical fibre, expanded by a 4× objective and then directed to a dichroic mirror ($\lambda > 700$ nm) reflecting it to an objective (100×, numerical aperture NA = 0.9, air) focusing on the sample. The sample was mounted on a remotely controlled three-axis piezo-electric stage, allowing positioning in a 100×100 µm$^2$ area range with nanometric accuracy. The induced luminescence from the sample was collected together with the scattered excitation beam by the same focusing objective; the dichroic mirror and a subsequent set of filters ($\lambda > 730$ nm) allowed a suitable ($> 10^{12}$) attenuation of the $\lambda = 690$ nm excitation component of the beam. The filtered beam was then focused with an achromatic doublet ($f = 100$ mm) and coupled into a graded-index multimode optical fibre ($\varnothing = 62.5$ µm) which takes the role of the pinhole aperture of the confocal system, other than guiding light to the detectors. The fibre was connected to an integrated 50:50 beam-splitter (BS) whose outputs led to Si based single-photon-avalanche photo-diodes (SPADs) operating in Geiger mode and counting electronics. The described configuration reproduces a "Hanbury Brown and Twiss" (HBT) interferometer [10], that allows the characterization of a single-photon emitter by the measurement of the second-order autocorrelation function $g^{(2)}(t)$ using the coincidence counts between the two detectors.

### 3. Microfabrication techniques

FIB microfabrication was performed with a Quanta 3D FEG DualBeam$^{TM}$ FIB apparatus, equipped with a liquid Ga$^+$ ion source and a field emission electron source. The nominal minimum ion beam spot diameter is 7 nm and the maximum ion energy is 30 keV. The beam deflection system is equipped with a pattern generation (NPGS - NanoPattern Generation

System, J.C. Nabity Systems, version 9.0) allowing for the definition of arbitrary milling shapes by means of CAD encoding.

The registration and fabrication of the SIL was performed with the experimental procedure described in the following.

Firstly, the confocal NIR-PL microscopy was employed to identify a suitable sub-superficial NIR-emitting colour centre with emission properties compatible with what reported in [29]. Subsequently, micro-grooves with depth of few tens of nanometers were FIB-milled at the sample surface, in order to provide a set of reference markers to register the position of the NIR centre. As shown respectively in Figs. 3a and 3b, the reference markers are clearly visible in confocal PL maps obtained by focusing on the sample surface. At a focal depth of ~6 µm below the sample surface the reference markers disappear and the map displays only the isolated NIR-emitting centre highlighted by the circle. After FIB fabrication, the reference markers are visible in PL because the milling process produces a thin layer of amorphous carbon at the sample surface, which exhibits a higher luminescence with respect to the surrounding background emission of the diamond crystal.

The depth position of the colour centre under investigation (~15 µm) was determined by multiplying the focal depth from the sample surface at which the emission rate from the centre is maximized (i.e. ~6 µm) by the diamond's refractive index ($n_d = 2.41$).

By changing the focal depth of the confocal PL microscopy apparatus, it is therefore possible to determine the position of the centre with respect to the markers with spatial resolution better than 1 µm, while the accuracy in the depth direction is lower (i.e. 2–3 µm), due to spherical aberrations.

With this registering procedure, a SIL was FIB-milled directly above the colour centre, as schematically shown in Fig. 4. The structure was obtained by drawing ~200 concentric circles at increasing fluence with a 30 keV Ga$^+$ ion current of 3 nA; the pattern was repeated several hundreds of times to achieve a smooth profile. A trench was milled around the lens (~50 circles), in order to allow for all rays within the desired acceptance angle to be extracted from the crystal (see Fig. 4). The fluence delivered at each 50×50 nm$^2$ pixel was determined by the dwell time. In first approximation, the milling rate was assumed to be directly proportional to the ion fluence, and consequently the desired milling pattern was converted into a corresponding dwell time pattern, for a specific ion current. During the FIB milling, the progressive shaping of the SIL structure was frequently imaged by SEM microscopy, in order to constantly monitor the fabrication process. In the above-mentioned experimental conditions, typical fabrication times for this kind of structure varied from 30 to 60 minutes.

As the purpose of the SIL fabrication is to fill the objective's numerical aperture (NA), the fabrication of a complete hemisphere was not necessary. On the other hand, a further reduction of the NA of the SIL would result in a decreasing light collection efficiency, but also in shorter fabrication times. A SIL of given $NA$ requires a maximum milling depth $D$ given by:

$$D = \left(1 - \sqrt{1 - NA^2}\right) d \tag{3}$$

where $d$ is the distance between the emitter and the sample surface (see Fig. 4).

In our case, $d = 15$ μm and a value $NA = 0.7$ is obtained for $D = 0.29 \cdot d = 4.4$ μm. The SIL diameter is $2d \cdot NA = 21$ μm.

Given these considerations, the best conditions for a rapid fabrication of an effective lens are met when the colour centre lies close (i.e. 2–3 μm) to the surface, because in this case it is possible to realize smaller SILs with greater NA. On the other hand, at such depth a higher accuracy in the determination of the depth position of the emitter is required.

After FIB microfabrication, the sample was cleaned in order to remove re-deposited amorphous material around the microstructure, which is detrimental both in terms of background luminescence and self-absorption. To this scope, the sample was annealed in vacuum at a temperature of 900 °C to graphitize the amorphized regions and to promote the segregation of implanted $Ga^+$ ions [35]. Electrochemical etching in de-ionized water (contact-less platinum electrodes, 200 V polarization, 30 mA current, few hours [36]) was then employed to selectively remove the graphitic phases. Subsequently, a final cleaning procedure in oxygen plasma (same parameters as described above) was performed.

## 3. Results and Conclusions

In Fig. 5a a SEM micrograph of the fabricated SIL structure is reported. Its smooth and regular shape can be appreciated, while in the surrounding regions the superficial marker structures are visible. In Fig. 5b a confocal PL map of the region where the SIL was fabricated is reported. The single NIR emitter is clearly visible as a bright isolated spot in the centre of the SIL, whose outer trench is distinguishable as a darker circular halo.

The increase in light collection efficiency is very significant. Before the fabrication of the SIL, the emission rate from the colour centre was $1.8\times10^4$ counts $s^{-1}$ in saturation conditions; it is worth noting that such a low count rate with respect to what is reported in literature is due to spherical aberrations, since the emitter is located ~15 μm below the sample surface, i.e. significantly deeper than in other reports. After the SIL fabrication, the emission rate reached $8.8\times10^4$ counts s$^{-1}$, i.e. a five-fold increase, thus demonstrating the effectiveness of the microstructure in increasing the photon collection efficiency.

As shown in Fig. 5b, the luminescent spot does not appear perfectly symmetric, indicating an astigmatic behaviour of the SIL. Possible explanations are a non-perfect centering of the SIL with respect to the emitter position and/or an imperfect spherical shape of the lens.

As more progress can be done in refining the fabrication strategy (improving the accuracy in the centre/lens alignment and in the lens shaping, increasing the numerical aperture, etc.), the combination of recently-discovered NIR-emitting centres in diamond and integrated solid immersion lenses demonstrated in this work represents a promising system to achieve the best single-photon emission performances (emission rate, stability, spectral properties) with respect to the state of the art, with appealing perspectives in the fields of single-photon emitters and quantum optics.

**Acknowledgments**

This research activity was funded by the following projects, which are gratefully acknowledged: FIRB "Future in Research 2010" project (CUP code: D11J11000450001) funded by the Italian Ministry for Teaching, University and Research (MIUR); EMRP project "EXL02-SIQUTE" (jointly funded by the EMRP participating countries within EURAMET and the European Union); "A.Di.N-Tech." project (CUP code: D15E13000130003) funded by University of Torino and Compagnia di San Paolo in the framework of the "Progetti di ricerca di Ateneo 2012" scheme; NATO SPS Project 984397; "Compagnia di San Paolo" project "Beyond classical limits in measurements by means of quantum correlations".

**Figure captions**

Fig. 1: Cone of light emerging from a colour centre positioned in the bulk of the pristine diamond sample (a) and of the diamond sample after SIL microfabrication (b).

Fig. 2: Schematic representation of the single-photon-sensitive confocal PL microscopy setup adopted in the present work.

Fig. 3: Confocal PL microscopy maps of the same region taken from different focal depths from the sample after the FIB fabrication of the reference markers. In (a) the focal depth coincides with the sample surface, and the reference markers are clearly visible. In (b) the focal depth is located ~15 μm below the sample surface: the reference markers are not visible while the single NIR-emitting centre highlighted by the circle is clearly visible.

Fig. 4: Schematic cross-sectional representation of the SIL geometry. All relevant distances are reported in units of the surface-emitter distance $d$.

Fig. 5: (a) SEM micrograph of the FIB-microfabricated SIL; the registration markers are visible in the surrounding region. (b) Confocal PL map of the region where the SIL was fabricated; the bright dot at the centre corresponds to the NIR emitter.

**Figures**

Fig. 1

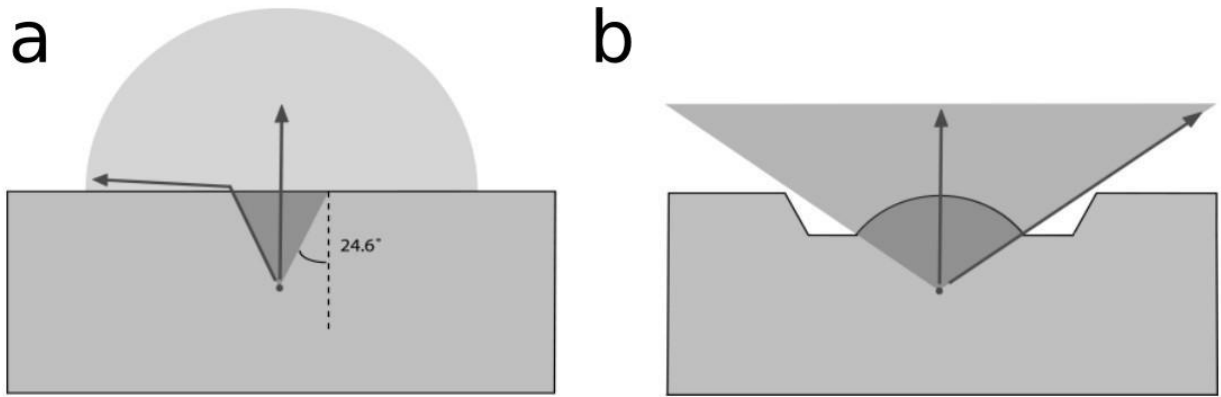

Fig. 2

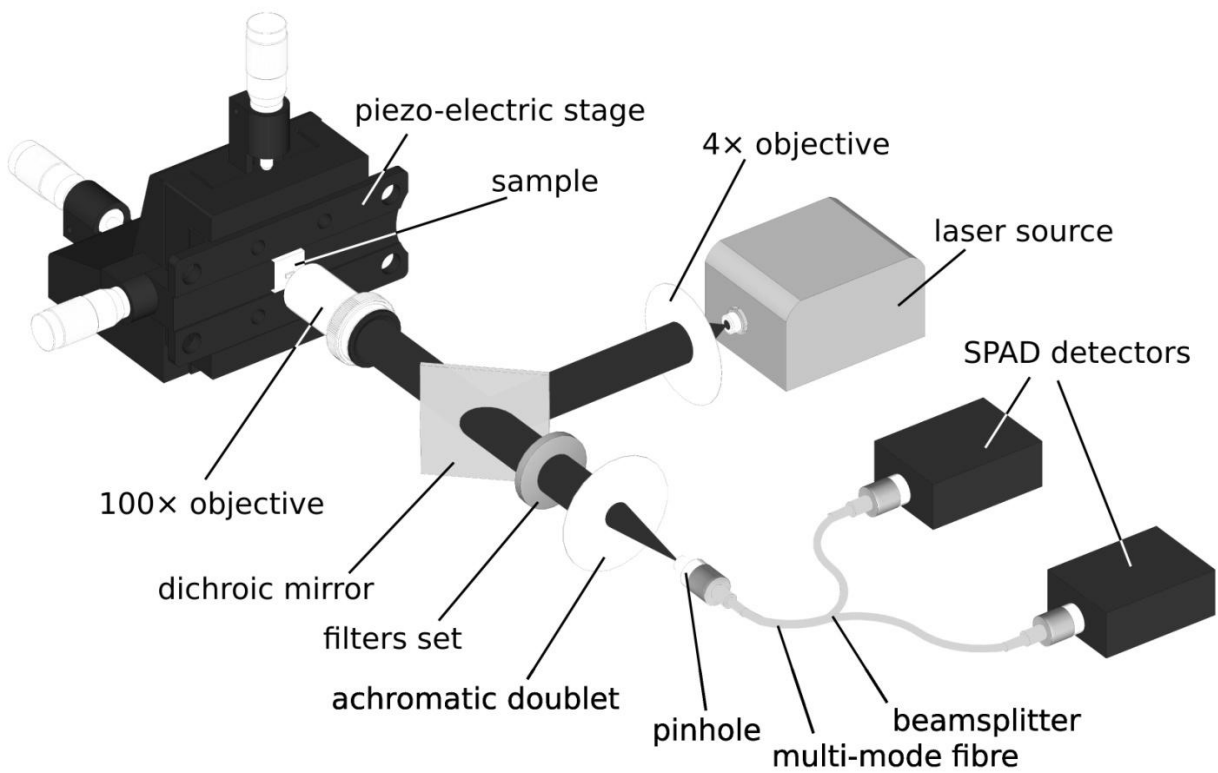

Fig. 3

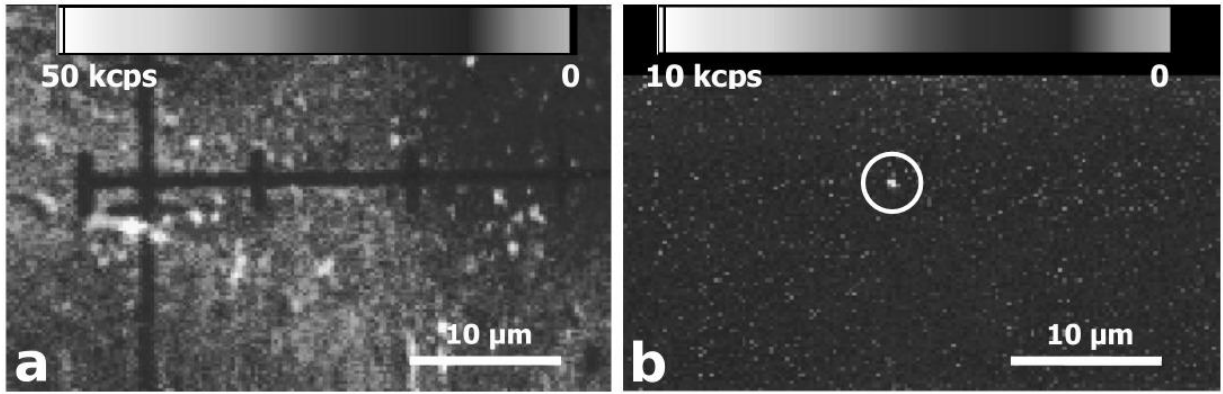

Fig. 4

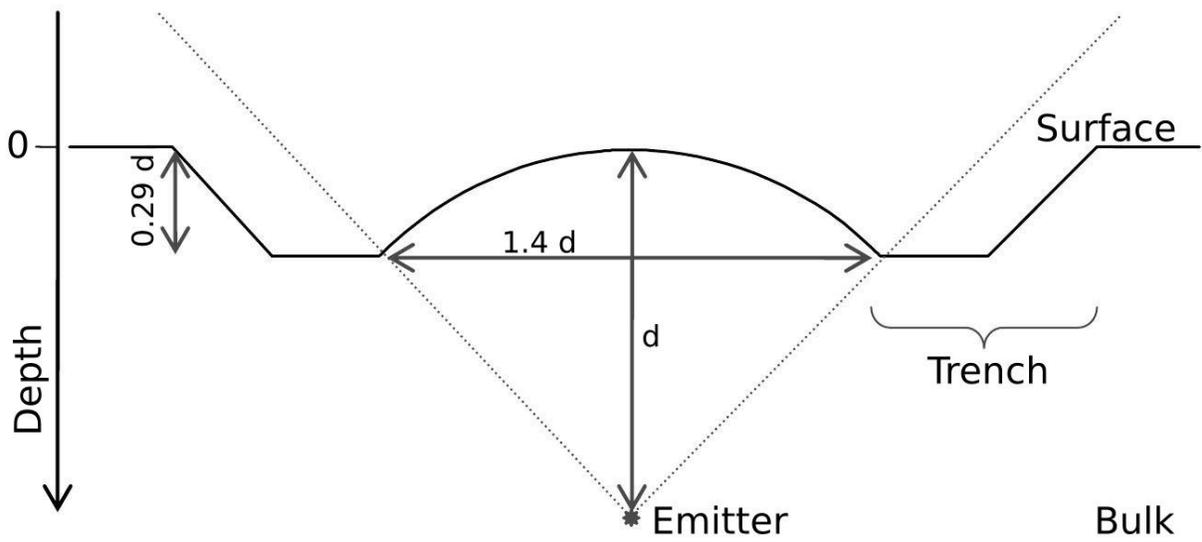

Fig. 5

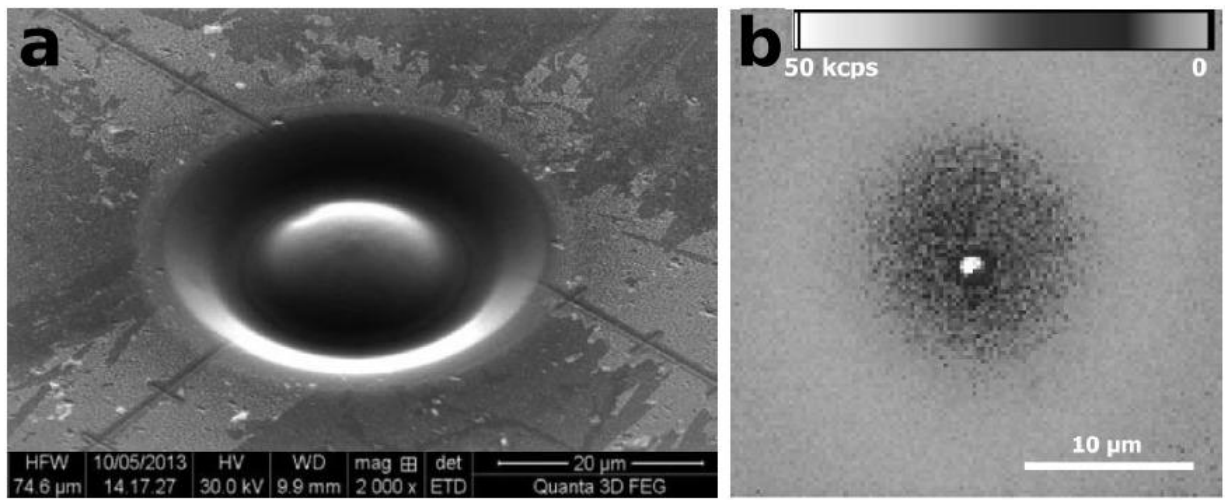